\documentclass{article}

\usepackage[preprint, nonatbib]{neurips_2024}

\usepackage{natbib}
\usepackage[utf8]{inputenc} 
\usepackage[T1]{fontenc}    
\usepackage{hyperref}       
\usepackage{url}            
\usepackage{booktabs}       
\usepackage{amsfonts}       
\usepackage{nicefrac}       
\usepackage{microtype}      
\usepackage{xcolor}         
\usepackage{graphicx}
\usepackage{algorithm}
\usepackage{algorithmic}
\usepackage{amsmath}
\usepackage{amssymb}
\usepackage{multirow}

\title{Large Language Models are Few-shot Generators: Proposing Hybrid Prompt Algorithm To Generate Webshell Escape Samples}

\author{%
  Mingrui Ma\thanks{Personal Webpage: \href{https://jkpathfinder.github.io}{https://jkpathfinder.github.io}} \\
  School of Cyber Science and Engineering\\
  Huazhong University of \\Science and Technology\\
  Wuhan, Hubei, China 430074 \\
  \texttt{m202271767@hust.edu.cn}
  \And
  Lansheng Han \\
  School of Cyber Science and Engineering\\
  Huazhong University of \\Science and Technology\\
  Wuhan, Hubei, China 430074 \\
  \texttt{1998010309@hust.edu.cn}
  \AND
  Chunjie Zhou \\
  The Key Laboratory of Ministry of Education for Image Processing and Intelligent Control\\
  School of Artificial Intelligence and Automation\\
  Huazhong University of Science and Technology\\
  Wuhan, Hubei, China 430074 \\
  \texttt{cjiezhou@hust.edu.cn}
}

\begin{document}

\maketitle

\begin{abstract}
  The frequent occurrence of cyber-attacks has made webshell attacks and defense gradually become a research hotspot in the field of network security. However, the lack of publicly available benchmark datasets and the over-reliance on manually defined rules for webshell escape sample generation have slowed down the progress of research related to webshell escape sample generation and artificial intelligence (AI)-based webshell detection. To address the drawbacks of weak webshell sample escape capabilities, the lack of webshell datasets with complex malicious features, and to promote the development of webshell detection, we propose the Hybrid Prompt algorithm for webshell escape sample generation with the help of large language models. As a prompt algorithm specifically developed for webshell sample generation, the Hybrid Prompt algorithm not only combines various prompt ideas including Chain of Thought, Tree of Thought, but also incorporates various components such as webshell hierarchical module and few-shot example to facilitate the LLM in learning and reasoning webshell escape strategies. Experimental results show that the Hybrid Prompt algorithm can work with multiple LLMs with excellent code reasoning ability to generate high-quality webshell samples with high Escape Rate (\(88.61\%\) with GPT-4 model on VirusTotal detection engine) and Survival Rate (\(54.98\%\) with GPT-4 model).
\end{abstract}

\section{Introduction} \label{Intro}

Webshell \cite{RF1}, as a typical example of malicious scripts exploiting injection vulnerabilities, allows hackers to remotely access and invade web servers, posing serious threats to network security. Similar to the research on malware detection, webshell generation, and detection are non-stationary, adversarial problems \cite{RF2}, which have been engaged in a constant game of cat and mouse, with an escalating spiral trend. From the attacker's perspective, mainstream webshell detection tools and engines like VirusTotal \cite{RF3}, WEBDIR+, and SHELLPUB are frequently updated and maintained, incorporating the rules and characteristics of new webshells within days or even shorter periods. This forces attackers to constantly develop new webshell generation methods to bypass the detection of such engines. On the detection side, research is still in its infancy \cite{RF4}. There is a lack of publicly available benchmark datasets and open-source baseline methods for webshell detection. Most models using neural networks or intelligent algorithms claim to have high accuracy and low false positives. However, the fact is that these models are tested on private datasets, which usually consist of only a few hundred or fewer samples, with obvious malicious features. In a real cyber-attack environment, the authenticity and generalization ability of such methods are difficult to guarantee. 

In fact, Abdelhakim et al. \cite{RF4} argued that AI methods excel at extracting abstract features in webshell, which are advanced features that go beyond lexical, syntactical, and semantical features. These advanced features help reveal hidden aspects in webshells that cannot be detected through syntax and semantic analysis. However, unlike the research on malware adversarial sample generation \cite{RF2, RF5, RF6, RF7}, research on webshell escape sample generation is still a blank field, which is due to the fact that the existing webshell bypass strategies are numerous and complicated, and there is no specific systematic method to follow. Therefore, it is an urgent and highly significant work to propose a webshell escape sample generation algorithm and construct a corresponding webshell benchmark dataset.

On the other hand, the blooming development of large language model (LLM) and artificial intelligence generated content (AIGC) technologies \cite{RF8} has already made an indelible impact in various domains such as chat and image generation \cite{RF9}. As the latest achievement in the field of natural language processing (NLP), LLM has taken a significant lead over earlier neural network structures (i.e. Long Short-Term Memory \cite{RF10}, Gate Recurrent Unit \cite{RF11}, etc.) in contextual reasoning and semantic understanding capabilities. The widespread application of LLM in various code-related tasks \cite{RF12, RF13, RF14, RF15, RF16, RF17} has fully showcased its excellent code reasoning abilities, making it possible to utilize LLM for generating webshell escape samples. Prompt engineering \cite{RF18} plays a crucial role in the vertical research application of LLM, which aims to explore better ways of human interaction with LLMs to fully leverage their performance potential. It is undeniable that many key techniques in prompt engineering, such as Chain of Thoughts (CoT) \cite{RF19}, Tree of Thoughts (ToT) \cite{RF20}, Zero-Shot CoT \cite{RF21}, etc., have improved the reasoning abilities of LLMs. Novel studies in prompt engineering, such as prompt finetuning, have been able to fine-tune the parameters in LLM, thus simplifying the traditional fine-tune process \cite{RF23}.

Therefore, in this work, we explore the unexplored research area of AIGC-enabled webshell escape sample generation strategies. We propose Hybrid Prompt, a hierarchical and modular prompt generation algorithm, and apply it to different LLM models to generate multiple webshell samples with high escape capabilities. Experimental results demonstrate that the escape samples generated by the Hybrid Prompt algorithm \(+\) various LLM models can bypass mainstream detection engines with high Escape Rate (\(ER\)) and Survival Rate (\(SR\)).

The main contributions of this paper are three-folds: 1) We propose the Hybrid Prompt algorithm, which combines the advantages of multiple prompt schemes such as ToT \cite{RF20}, few-shot prompting, CoT, etc. By synthesizing key features related to webshell escape and designing prompt strategies tailored to different sizes of webshells, the algorithm effectively enhances the code reasoning ability of LLM models and generates high-quality webshell escape samples; 2) We construct a webshell benchmark dataset generated by the Hybrid Prompt algorithm. This dataset achieves high \(ER\) and \(SR\) among mainstream detection engines and reflects the performance of rule-based detection engines more realistically and effectively; 3) We investigate and compare the quality of escape samples generated by different LLM models using the Hybrid Prompt algorithm. All these samples exhibit high \(ER\), surpassing webshell samples generated by other intelligent algorithms (i.e. genetic algorithm \cite{RF24}). See App. \ref{sec:appendix Virustotal detect} for further detailed examples and preliminaries.

\section{Algorithm design} \label{2}

\subsection{Overall workflow}

\begin{figure}[htbp]
\centerline{\includegraphics[width=0.85\textwidth]{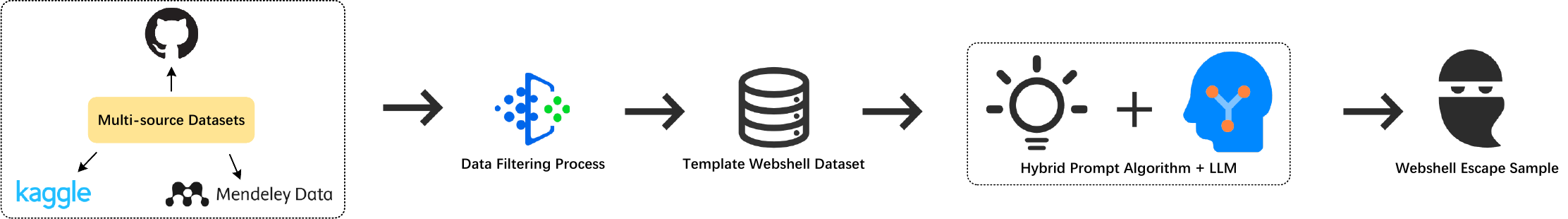}}
\caption{The overall workflow of webshell escape sample generation.}
\label{Fig4}
\end{figure}

The overall flow from collecting multi-source webshell scripts to generating webshell escape samples is shown in Figure \ref{Fig4}.

Specifically, we first collect 28,770 raw samples with obfuscated formats from mainstream data websites such as GitHub and Kaggle. Data filtering is then performed to obtain a clean Template webshell dataset with 6,639 samples. Through the Hybrid Prompt algorithm, LLM engages in self-inference based on the template webshell code and generates high-quality webshell escape samples.

\subsection{Data filtering} \label{3.2}

\begin{figure}[htbp]
\centerline{\includegraphics[width=0.6\textwidth]{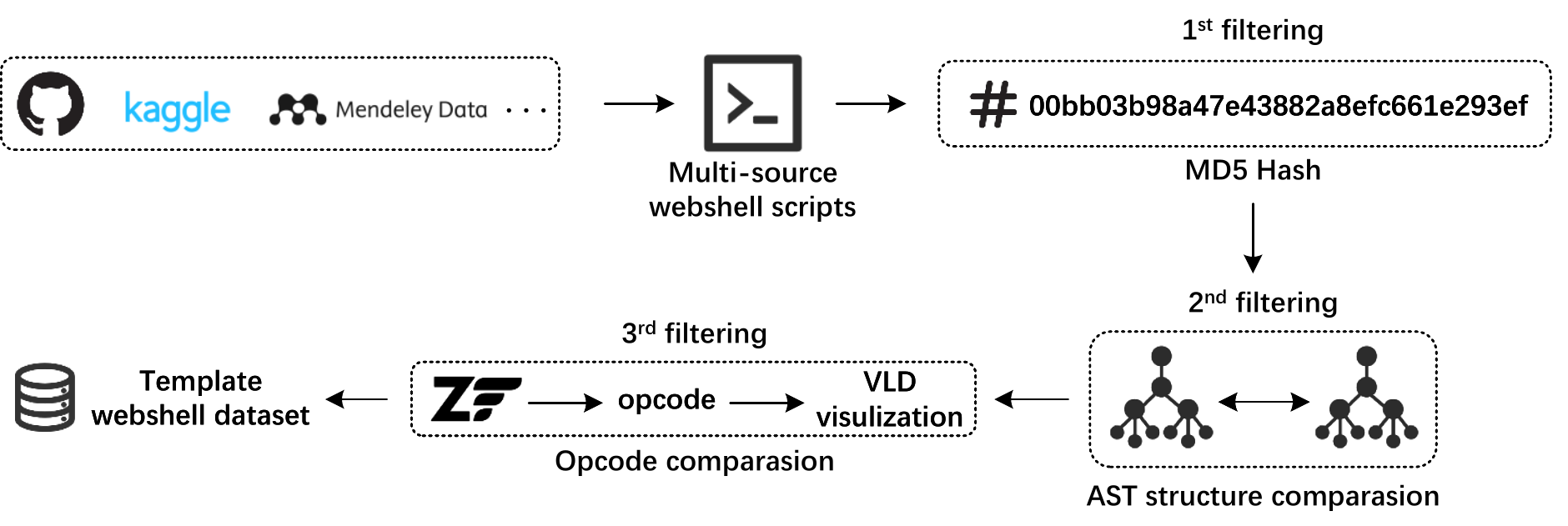}}
\caption{Triple data filtering process.}
\label{Fig5}
\end{figure}

To enhance LLM's learning and understanding of the input webshell samples, we need to construct the Template webshell dataset, a clean and well-characterized webshell dataset with file names unified by \textit{MD5} hash values. We perform triple data filtering process on multi-source webshell datasets, as shown in Figure \ref{Fig5}.

In the first filtering step, we calculate the \textit{MD5} hash value of all scripts to filter out webshell scripts with consistent content but confusing names. In the second filtering step, we convert the webshell scripts into Abstract Syntax Tree (AST) structures to filter out the scripts with the same syntax structure. For PHP scripts, we use "\textit{php-ast}" to perform the translation (\textit{ast\textbackslash parse\_code}) and add the \textit{name, kind} attribute to the nodes. The pseudocode for this step is shown in Algorithm \ref{alg:php-ast}.

\begin{algorithm} \footnotesize
	\caption{Php-ast Runtime Flow}
    \label{alg:php-ast}
	\begin{algorithmic}[1]
		\STATE {\$ast = ast\textbackslash parse\_code(\$code, \$version=70);}
        \STATE {\$new\_ast = add\_attr(\$ast);}
        \STATE {\$json = json\_encode(\$new\_ast, JSON\_PRETTY\_PRINT \(\vert\) JSON\_UNESCAPED\_UNICODE \(\vert\) JSON\_OBJECT\_AS \_ARRAY);}
	\end{algorithmic}
\end{algorithm}

In the third filtering step, the Vulcan Logic Disassembler (VLD) module in Zend engine is used to disassemble the scripts into opcode structures, aiming to filter out webshell scripts with consistent execution sequences. See details in App. \ref{sec:appendix AST and VLD}.

\subsection{Hybrid Prompt}

\begin{figure}[htbp]
\centerline{\includegraphics[width=0.95\textwidth]{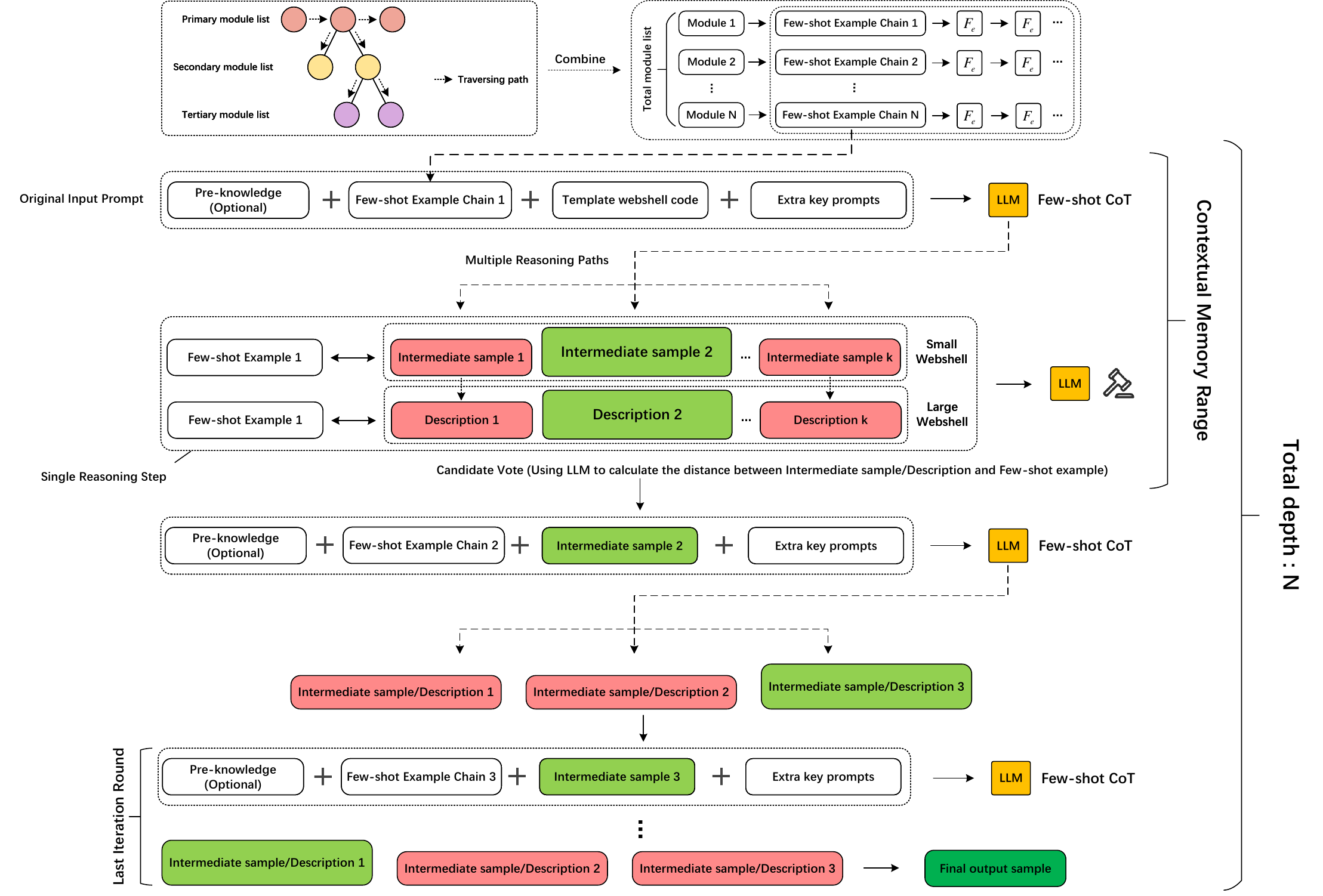}}
\caption{The flowchart of Hybrid Prompt algorithm.}
\label{Fig8}
\end{figure}

The ToT method has performance advantages over CoT, Self Consistency (SC) \cite{RF28} method in solving complex reasoning problems by searching for multiple solution paths, using strategies such as backtracking and pruning. However, this is not yet sufficient to address the heuristic search task of generating webshell escape samples with a broader search space and stricter normalization constraints. Therefore, we propose the Hybrid Prompt algorithm, whose overall flow is shown in Figure \ref{Fig8}.

We begin with the normalized definition of relevant symbols. We use \(M\) to denote LLM, \(o\) to denote one of the candidates generated by each module of Hybrid Prompt, \(O\) to denote the set composed of candidates, \(x\) to denote the original input of Hybrid Prompt, \(F_e\) to denote the few-shot example, \(N\) to denote the tree depth of Hybrid Prompt and \(p\) to denote the number of candidates.

\subsubsection{Module setting}

To divide the task of generating webshell escape samples into several subtasks and enhance LLM's single-step reasoning ability in the Hybrid Prompt algorithm, we introduce the concept of modules. A module represents a key escape idea, such as "Add Unrelated Comments", and some modules further have secondary and tertiary modules, see details in App. \ref{sec:appendix Hms}. Each module contains an \(F_e\) chain consisting of several \(F_e\) nodes, which will be further elaborated in Section \ref{Compound Input Generator}. Therefore, in Hybrid Prompt, LLM's single-step search space contemplates the escape plan guided by a module for the current round of input webshell.

\subsubsection{Compound input generator \(G(M,o)\)} \label{Compound Input Generator} 

To alleviate the issue of hallucination during LLM's single-step reasoning, which may result in generating corrupted webshells, we apply the CoT method to generate multiple intermediate webshell samples at each reasoning step (each module) in the Hybrid Prompt algorithm, as illustrated in Figure \ref{Fig8}. Considering that LLM may generate some low-value solutions with large deviations from the expectation, thus reducing the efficiency of subsequent votes, we design \(F_e\) chain structure for each module, see details in App. \ref{sec:appendix Fechain}. Therefore, \(G(M, o) = M(F_e, o)\). When filtering the \(F_e\) chain, we adhere to the following principles: 1) The structure of the example webshell code should be as simple as possible; 2) Each node contains, as far as possible, only the processing methods corresponding to that module. The purpose is to reduce the difficulty of LLM in learning the corresponding method through an example that is as simple as possible and contains the core idea. The descriptive explanation further enhances the interpretability of the solutions. This idea is also in line with the logical process of human learning and cognition, to help LLM better learn the features of the methods.

\(F_e\) can essentially "modify" the LLM’s thinking direction to a certain extent so that webshell can be generated in a Few-shot CoT mindset. In most cases, each \(F_e\) chain contains multiple \(F_e\) examples to provide more comprehensive coverage of different scenarios. In this case, multiple nodes are used as input prompt components for the current iteration round, to help LLM better learn multiple segmented strategies. Due to the large search space and sample diversity for each module, this Few-shot CoT method yields better results. Therefore, in each reasoning step of the Hybrid Prompt algorithm, the compound input consists of Pre-knowledge (optional), \(F_e\) chain, input webshell code, and extra key prompts, as shown in Figure \ref{Fig8}.

Meanwhile, based on the input webshell size, we design 2 different generation approaches. For small webshells, we include \(p\) candidate webshell samples in a single conversation returned by the LLM. In this case, the average maximum length of each candidate webshell sample  \(L(Avg\_Candidate_i)\) is calculated as \(L(Avg\_Candidate_i)=(L(MaxToken) - L(InputPrompt))/p.\) Where \(L(MaxToken)\) denotes the maximum context length that the current LLM model can handle, and \(L(InputPrompt)\) denotes the length of the input prompt in the current thought. Since small webshells are generally shorter, this approach can save the consumption of LLM’s token resources, and enable LLM to generate more diverse samples in the returned message of a single conversation through specific "key prompts".

For large webshells, we enable the \(n\) parameter function to generate \(p\) candidate webshell samples by receiving multiple return messages from LLM. In this case, the maximum length of each candidate webshell sample \(L(Candidate_i)\) is calculated as \(L(Candidate_i) = L(MaxToken) - L(InputPrompt) - L(Description_i).\) Where \(Description_i\) represents the brief description generated by LLM for the \(i^{th}\) candidate webshell sample, which is used to summarize the idea of candidate webshell generation and facilitate the subsequent voting process. This approach maximizes the length of the generated candidate webshell sample at the expense of consuming more token resources.

\subsubsection{Candidate evaluator \(V(M,O)\)}

The candidate evaluator is also designed to have 2 different voting methods for large and small webshells. For small webshells, Hybrid Prompt uses LLM to vote on multiple intermediate webshell samples and filter out the optimal ones. The reason for voting on multiple samples instead of voting on solutions is two-fold: 1) Since the compound input generator operates in a few-shot CoT mindset, webshell samples help LLM evaluate and assess the differences between generated examples more intuitively to make optimal judgments; 2) Voting directly on the samples can preserve all the original information of the candidate webshells. In this case, \(L(Generator(Input+Output)) \approx L(Evaluator(Input+Output)) \textless L(MaxToken).\) Because both contain \(F_e\)s, the webshell contents of \(p\) candidates, and additional prompt information.

For large webshells, it is not feasible to directly input the webshell contents of \(p\) candidates into LLM because \(p \times (L(Candidate_i)) + L(F_e chain) + L(Additional_{Prompt}) \textgreater L(MaxToken).\) Therefore, we use \(Description_i\) instead of \(Candidate_i\) as the input component of the voting procedure. This kind of information compression idea will inevitably lose the original code information. App. \ref{sec:appendix vote} presents a specific example comparing 2 voting ideas.

Regardless of the voting idea, for \(V(M,O)\), where \(O=\{o_1,o_2,...,o_p\}\), \(V(M,o_i)=1\) is considered a good state, when \(o_i \sim M^{vote} (o_i | O).\) For Hybrid Prompt, the evaluation of a good state is to synthesize both the confusion level of the intermediate results generated by LLM for a module and the distance between them and the \(F_e\)s, as shown in Figure \ref{Fig8}. By allowing LLM to pursue local optimal solutions at each step of sample generation, this "greedy" idea makes it easier for the LLM to approximate the global optimal solution for the heuristic problem of escape sample generation.

\subsubsection{Search algorithm} \hfill

For the Hybrid Prompt method, the depth of the tree \(N\) depends on the number of the modules selected. The DFS strategy leads to an excessive state space of LLM during the backtracking and pruning stages, which reduces the efficiency of the algorithm operation. Therefore, we consider using the BFS search algorithm. The pseudocode of the corresponding Hybrid Prompt-BFS algorithm is shown in Algorithm \ref{alg:BFS}. The final output of the webshell escape sample is the candidate that wins in the vote process at the \(N^{th}\) layer.

\begin{algorithm} \footnotesize
	\caption{Hybrid Prompt-BFS Algorithm}
    \label{alg:BFS}
	\begin{algorithmic}[1]
		\REQUIRE Input \(x\), Compound Input Generator \(G(M,o)\), Candidate Evaluator \(V(M,O)\), Tree Depth \(N\), Candidate num \(p\), Step Output \(O_i (O \leq i \leq N)\)          
		\STATE \(O_0 = x\)
        \FOR{ \(n\) = 1 to \(N\) }
            \STATE \(O_n^{'} = \{ [o,z] \vert o \in O_{n-1}, z_n \in G(M,o) \}\)
            \STATE \(V_n = V(M, O_n^{'})\)
            \STATE \(O_n = sort (V_n, p)\)
        \ENDFOR    
        \STATE Return \(O_n\)
	\end{algorithmic}
\end{algorithm}

\subsubsection{Additional explanation} \hfill

Since LLM has a limited range of contextual memory, we cannot let LLM memorize the entire Hybrid Prompt context but should set its local memory range. For this reason, our approach is to set the contextual memory range for the Hybrid Prompt, as shown in Figure \ref{Fig8}. For the webshell escape sample generation task, an important guiding principle is to ensure the validity of generated samples. This means that the escaped samples should not lose the attack behavior and malicious features of the original samples and can be executed correctly without any syntax or lexical errors. To achieve this, Hybrid Prompt introduces Safeguard Prompt to constrain sample generation and improve \(SR\). In addition, common techniques in prompt engineering, such as ``` delimiter, are also applied in the Hybrid Prompt algorithm to normalize the output of LLMs. The order of modules also has a significant impact on the Hybrid Prompt algorithm. Therefore, when running the Hybrid Prompt algorithm, it is important to consider the relative position between specific modules and establish corresponding rules to avoid such situations from occurring. See more explanations in App. \ref{sec:appendix CMR}.

\section{Experiments} \label{Experiments}

In the experimental section, our main objective is to answer the following questions:

\textbf{RQ1: Effectiveness analysis.} Can Hybrid Prompt effectively generate escape samples? \textbf{RQ2: Ablation study.} Are the individual parts of the Hybrid Prompt algorithm effectively designed? \textbf{RQ3: Sensitivity analysis.} Does the number of candidates \(p\) affect the performance of the Hybrid Prompt algorithm? What is the impact on the performance of the Hybrid Prompt algorithm if the tree depth \(N\) is reduced by half? \textbf{RQ4: Defensive study.} Is there a method available to detect the escape samples generated by the Hybrid Prompt algorithm?

\subsection{Setup} \label{3.1}

\textbf{Experimental settings.} We use the Hybrid Prompt algorithm with the tree depth \(N = 10\) to generate a total of 3,273 escape samples from 6,639 samples in the Template webshell dataset, with 1,091 samples generated by GPT-3.5, GPT-4, and Code-llama-34B each. We implement Hybrid Prompt using Python 3.10 and PyTorch 1.8.1. We run all our experiments on a Windows server configured with a 2.9GHz Intel Xeon 6326R CPU, 2 \(\times\) 80G NVIDIA TESLA A100 GPU, and 64GB memory. We further build a Virtual Attack Environment for verifying the validity of the escape samples generated by the Hybrid Prompt algorithm.

\textbf{Evaluation metrics \& Comparative methods.} To better compare the quality of samples generated by different LLM models using the Hybrid Prompt algorithm, we choose 3 evaluation metrics: \(ER\), \(SR\) and Modification Ratio (\(MR\)), where \(ER = 1-DR\), \(DR = N_{Detected\_samples} / N_{Total\_samples}\), \(SR = N_{Malicious\_samples} / N_{Total\_samples}\). \(DR\) represents the detection accuracy, \(N_{Total\_samples}\) is the total number of samples generated by LLM under the Hybrid Prompt algorithm, and \(N_{Malicious\_samples}\) is the number of samples generated by LLM under the Hybrid Prompt algorithm that still retain malicious functionality. \(MR = N_{Escape\_samp\_size} / N_{Ori\_samp\_size}\), \(N_{Escape\_samp\_size}\) represents the total file size of \(N\) escape samples generated, \(N_{Ori\_samp\_size}\) represents the total file size of \(N\) original samples generated. Due to the lack of relevant research, we also include a comparison with the dataset from CWSOGG \cite{RF24}, an obfuscated webshell dataset generated using the genetic algorithm.

\textbf{Models \& Detection engines.} We test the \(ER\), \(SR\), and \(MR\) of samples generated by Hybrid Prompt under 4 detection engines: Web Shell Detector, WEBDIR+, SHELLPUB and VirusTotal respectively. In addition, we cross-check the performance of several LLM models, including GPT-3.5, GPT-4, and Code-llama-34B, which demonstrate excellent performance in code generation and semantic understanding tasks. For VirusTotal, we set the Label Aggregation threshold to 13 following the tuning recommendation of Zhu et al. \cite{ZhuShuo} and apply redundant votes to certain vulnerable detection engines (such as Comodo). 

For further details on the Experimental setup, see App. \ref{sec:appendix Vae}. Codes available at \href{https://github.com/HybridPrompt/Hybrid-Prompt-demo}{https://github.com/HybridPrompt/Hybrid-Prompt-demo}

\subsection{Comparative experiment}

To answer \textbf{RQ1}, the comparative results are shown in Table \ref{Tab2}.

\begin{table}[htbp]\scriptsize
\caption{Comparative experiment results}
\label{Tab2}
\centering
\begin{tabular}{ccccccc}
\toprule
Anti-Virus Engine & \textit{Web Shell Detector} & \textit{WEBDIR+} & \textit{VirusTotal} & \textit{SHELLPUB} & \multirow{2}{*}{\(SR\)} & \multirow{2}{*}{\(MR\)} \\
\cmidrule(r){2-5} 
Model & \multicolumn{4}{c}{\(ER\)} \\
\midrule
GPT-3.5 Turbo + Hybrid Prompt & 0.9342 & 0.8874 & 0.7465 & 0.8023 & 0.4093 & 1.38 \\
GPT-4 + Hybrid Prompt & 0.9727 & 0.9287 & 0.8861 & 0.9024 & 0.5498 & 1.43 \\
Code-llama-34B + Hybrid Prompt & 0.9015 & 0.8549 & 0.6358 & 0.7625 & 0.3021 & 2.95 \\
Original Template Dataset & 0.3415 & 0.2054 & 0.1232 & 0.1684 & 1 & N/A \\
CWSOGG Dataset & 0.4052 & 0.3151 & 0.2327 & 0.2748 & 1 & N/A \\
\bottomrule
\end{tabular}
\end{table}

In Table \ref{Tab2}, the GPT-4 + Hybrid Prompt algorithm has the best comprehensive performance, leading to both \(ER\) and \(SR\). This is because GPT-4 is more capable of following complex instructions carefully, while Hybrid Prompt contains multiple detailed instructions with normalized constraints. GPT-3.5, on the other hand, could partially follow complex instructions, resulting in a higher probability of generating escape samples that prioritize either \(ER\) or \(SR\), making it difficult to balance both. It is encouraging to note that the comprehensive performance of the open-source LLM Code-llama-34B, is very close to that of the GPT-3.5 model, confirming the performance potential of the open-source models. Meanwhile, the \(ER\) of webshell samples generated by the 3 LLM models + Hybrid Prompt algorithm have far exceeded those of the Original template dataset and the CWSOGG dataset, which fully demonstrate the performance superiority and dominance of the LLM models over rule-based artificial escape strategies and the traditional intelligent algorithms (i.e., genetic algorithm). However, when considering the \(MR\) metric, the open-source LLM Code-llama-34B still lags behind GPT-3.5 and GPT-4, indicating the need to generate more obfuscated content to evade detection. As for the detection engines, VirusTotal, due to its integration of many different detection engines, has a higher overall \(DR\) compared to Web Shell Detector, SHELLPUB and WEBDIR+. However, even VirusTotal struggles with the creativity of LLMs and the uncertainty of the generated escape samples, which illustrates the limitations and drawbacks of these types of specific rule-based detection engines.

\subsection{Ablation analysis}

To validate the effectiveness of the algorithm and address \textbf{RQ2}, we test the performance of samples generated by removing different components of the algorithm (i.e. removing Safeguard Prompt, removing \(F_e\) chain, removing voting strategy, removing the entire Hybrid Prompt algorithm) under \(ER\) and \(SR\) evaluation metrics in the GPT-3.5 model. The experimental results are shown in Table \ref{Tab3}.

\begin{table}[htbp]\scriptsize
\caption{The comparative results of ablation analysis}
\label{Tab3}
\centering
\begin{tabular}{cccccc}
\toprule
Anti-Virus Engine & \textit{Web Shell Detector} & \textit{WEBDIR+} & \textit{VirusTotal} & \textit{SHELLPUB} & \multirow{2}{*}{\(SR\)}\\
\cmidrule(r){2-5} 
Strategy & \multicolumn{4}{c}{\(ER\)} \\
\midrule
Hybrid Prompt & 0.9342 & 0.8874 & 0.7465 & 0.8023 & 0.4093 \\
W/o Safeguard Prompt & 0.9221 & 0.8653 & 0.7114 & 0.7721 & 0.3398 \\
W/o \(F_e\) & 0.7315 & 0.6819 & 0.5042 & 0.6911 & 0.2310 \\
W/o Voting & 0.8213 & 0.7998 & 0.6524 & 0.7039 & 0.3067 \\
W/o Hybrid Prompt & 0.5021 & 0.4267 & 0.3120 & 0.3863 & 0.1513 \\
\bottomrule
\end{tabular}
\end{table}

Table \ref{Tab3} illustrates that "W/o Hybrid Prompt" has poor performance and a high probability of hallucination due to the absence of any additional prompt. Both "W/o \(F_e\)" and "W/o Voting" produce different degrees of performance degradation. For "W/o \(F_e\)", LLM loses reference examples, leading to a higher probability of generating corrupted samples. "W/o \(F_e\)" also indirectly reflects that the current LLM's code reasoning ability still relies on \(F_e\) chains to achieve better task performance. For "W/o Voting", LLM is unable to explore multiple reasoning paths, so the generation space and diversity of samples are limited, which leads to a lower \(ER\). "W/o Safeguard Prompt" has the least impact on the quality of generated escape samples. Although the probability of generating corrupted samples increases and the \(SR\) decreases due to the loss of Safeguard Prompt's normalization measures, the impact on the \(ER\) is not significant. However, all ablation models produce varying degrees of performance degradation compared to the complete Hybrid Prompt algorithm, fully demonstrating the effectiveness of various components in the Hybrid Prompt algorithm. 

\subsection{Sensitivity analysis} \label{3.4}
We investigate the impact of the candidate number, \(p\), and the tree depth, \(N\), on the \(SR\) and \(ER\) evaluation metrics of generated samples in the GPT-3.5 model. The experimental results of \textbf{RQ3} are shown in Table \ref{Tab4}.

\begin{table}[htbp]\scriptsize
\caption{The comparative results of sensitivity analysis}
\label{Tab4}
\centering
\begin{tabular}{cccccc}
\toprule
Anti-Virus Engine & \textit{Web Shell Detector} & \textit{WEBDIR+} & \textit{VirusTotal} & \textit{SHELLPUB} & \multirow{2}{*}{\(SR\)} \\
\cmidrule(r){2-5} 
Candidate num \(p\) & \multicolumn{4}{c}{\(ER\)} \\
\midrule
1 & 0.8213 & 0.7998 & 0.6524 & 0.7039 & 0.3067 \\
2 & 0.8749 & 0.8567 & 0.7031 & 0.7687 & 0.3648 \\
3 & 0.9342 & 0.8874 & 0.7465 & 0.8023 & 0.4093 \\
4 & 0.9489 & 0.8968 & 0.7621 & 0.8095 & 0.4163 \\
5 & 0.9522 & 0.9014 & 0.7708 & 0.8193 & 0.4266 \\
\toprule
Anti-Virus Engine & \textit{Web Shell Detector} & \textit{WEBDIR+} & \textit{VirusTotal} & \textit{SHELLPUB} & \multirow{2}{*}{\(SR\)} \\
\cmidrule(r){2-5}  
Tree depth \(N\) & \multicolumn{4}{c}{\(ER\)} \\
\midrule
5 & 0.8957 & 0.8210 & 0.5982 & 0.6724 & 0.4102 \\
10 & 0.9342 & 0.8874 & 0.7465 & 0.8023 & 0.4093 \\
\bottomrule
\end{tabular}
\end{table}

From Table \ref{Tab4}, it can be observed that a larger number of candidates can increase the search space of LLM, which in turn enriches the diversity of generated samples, enables better selection of the optimal solution, and improves the sample \(ER\) and \(SR\). However, the increase of \(p\) will also result in a higher token consumption and, in the case of small webshells, further reduces the \(L(Avg\_Candidate_i)\) for each sample. Table \ref{Tab4} also shows the "marginal effect" that occurs as \(p\) increases. When \(p\) exceeds 3, the performance improvement of \(ER\) and \(SR\) metrics is not obvious, which can be attributed to the fact that the search space of LLM's self-inference is approaching the local upper limit. However, it is noteworthy that the consumption of tokens exhibits an almost linear relationship with the increase in \(p\), despite the limited performance gains in \(ER\) and \(SR\) metrics. Therefore, the pros and cons between evaluation metrics and resource consumption should be weighed in practical applications. When the tree depth \(N = 5\), the \(ER\) of the Hybrid Prompt algorithm shows a significant decrease. From this, it is not difficult to analyze that Hybrid Prompt progressively enhances the level of obfuscation in candidate samples through the step-by-step inference in each module, aiming to reach the goal of high \(ER\). However, there is only a slight difference in \(SR\) between \(N = 5\) and \(N = 10\), indicating that a series of normalization measures of Hybrid Prompt algorithm effectively guarantees the validity of the generated samples. See App. \ref{sec:appendix Visualization} for visualization results of the experiments.

\subsection{Defensive study} \label{3.5}

As mentioned in Section \ref{Intro}, we aim to facilitate research on AI-based webshell detection algorithms rather than promote malicious activities. Considering the lack of available open-source solutions for AI-based webshell detection methods \cite{MaMing} and to inspire \textbf{RQ4}, we design a comparative AI-webshell detection engine based on the BERT \cite{BERT} model. In short, we use 2 BERT models with the identical structure (named Model 1 and Model 2). During the fine-tuning stage, Model 2 additionally introduces escape samples generated by the Hybrid Prompt algorithm. Both Model 1 and Model 2 are tested on the same dataset using \(Acc\) and \(F1\) as evaluation metrics where \(Acc = (TP+TN)/(TP+FP+TN+FN)\), \(F1= (2 \times Pre \times Rec)/(Pre + Rec)\), \(Pre = TP/(TP+FP)\), \(Rec = TP/(TP+FN)\). For specific methodology and experimental settings, please refer to App. \ref{sec:appendix Defensive}. The test results are shown in Table \ref{Tab5}.

\begin{table}[htbp]\scriptsize
\caption{Defensive study results}
\label{Tab5}
\centering
\begin{tabular}{cccc}
\toprule
\multicolumn{2}{c}{Model 1} & \multicolumn{2}{c}{Model 2} \\
\midrule
Acc & F1 & Acc & F1 \\
0.6917 & 0.8059 & 0.9939 & 0.9958 \\
\bottomrule
\end{tabular}
\end{table}

We can observe that the adversarial fine-tuned Model 2 outperforms the baseline fine-tuned Model 1 in both evaluation metrics, with \(30.22\%\) and \(18.99\%\) improvement in \(Acc\) and \(F1\) metrics, respectively. This is because the adversarial fine-tuned Model 2 effectively learns the abstract features of escape samples, which results in high-precision recognition of escape strategies during the testing phase. It is worth noting that the experimental results in Table \ref{Tab5} are obtained with an imbalanced data distribution. If the number of benign samples is close to the number of malicious samples, the performance gap between Model 2 and Model 1 will be further amplified. Table \ref{Tab5} also demonstrates that the webshell escape samples generated under the guidance of the Hybrid Prompt algorithm are valuable for enhancing the robustness and generalization ability of the AI-based webshell detection engines, and can effectively achieve the goal of promoting defense by attack.

\section{Related work}
\textbf{Prompt engineering algorithm.} As one of the most classic prompt algorithms, CoT \cite{RF19} aims to assist LLMs in achieving complex reasoning abilities through intermediate inference steps. Zero-shot CoT \cite{RF21}, as a follow-up to CoT, enables LLM to perform self-reasoning through twice generation, involving 2 separate prompting processes. SC \cite{RF28} serves as another complement to the CoT algorithm by sampling a diverse set of reasoning paths and marginalizing out reasoning paths to aggregate final answers. Least to Most Prompting (LtM) \cite{RF29}, also an advancement of the CoT algorithm, decomposes a problem into a set of subproblems built upon each other and inputs the solutions of the previous sub-problem into the prompt of the next sub-problem to gradually solve each sub-problem. Generated Knowledge Approach (GKA) \cite{RF30} enables LLM to generate potentially useful information related to a given question before generating the response through 2 intermediate steps: knowledge generation and knowledge integration. Diverse Verifier on Reasoning Steps (DiVeRSe) \cite{RF31}, on the other hand, improves the reliability of LLM answers by generating multiple reasoning paths. 

\textbf{The application of LLM in code-related tasks.} Zhang et al. \cite{RF32} utilized ChatGPT to generate vulnerability exploitation code. Liu et al. \cite{RF33} applied GPT to the task of vulnerability description mapping and evaluation tasks. They provided certain prompts to ChatGPT and extracted the required information from its responses using regular expressions. Zhang et al. \cite{RF34} proposed STEAM, a framework for bug fixing using LLM to simulate programmers' behaviors. Kang et al. introduced the LIBRO \cite{RF35} model for exploring bug reproduction tasks. The aforementioned researches demonstrate that with appropriate algorithmic design, LLM is capable of handling various specific tasks in the field of code analysis.

\textbf{Researches on webshell detection techniques.} We categorize the research in the field of webshell detection into 3 stages: Start Stage, Initial Development Stage, and In-depth Development Stage. In the Start Stage, research methods are simple and have numerous flaws and deficiencies, such as limited private datasets, unreasonable feature extraction methods \cite{RF36, RF37}, etc. In the Initial Development Stage, relevant studies explore and make progress in various aspects of the detection process. However, theoretical innovations remain relatively scarce \cite{RF38, RF39, RF40, RF41, RF42}, etc.  In the In-depth Development Stage, simple individual classifiers or machine learning algorithms become less common, and related research has penetrated into the theoretical process level of modeling methods \cite{RF43, RF44}. However, from an overall point of view, research related to webshell detection techniques is still in its early stages, largely due to the slow progress of the attacker's research, and the lack of advanced webshell escape sample generation algorithms in the field.

\section{Conclusion} \label{5}

In this paper, we propose Hybrid Prompt, a webshell escape sample generation prompt algorithm that combines various prompt strategies such as ToT, CoT, etc. Hybrid Prompt combines structured webshell module and \(F_e\) chain, utilizes auxiliary methods to inspire LLMs to perform self-assessment and optimization, and demonstrates excellent performance on LLMs with strong code reasoning capabilities (GPT-3.5, GPT-4, Code-llama-34B), enabling the generation of high-quality webshell escape samples. Hybrid Prompt algorithm also exhibits strong scalability and generalization capability, allowing for the addition of more modules and corresponding \(F_e\) chains to update escape strategies and expand to more webshell languages.

\paragraph{Limitations and future work.} The Hybrid Prompt algorithm currently supports a limited number of webshell languages, and there is a need to expand it to support more webshell languages in the future. Hybrid Prompt algorithm does not fine-tune LLMs. Fine-tuning can further reduce the probability of LLM hallucination and improve the quality of generated escape samples. For the voting strategy in the case of large webshells, the description-based strategy used in the Hybrid Prompt algorithm results in the loss of original information from candidate code, which in turn affects the vote effect of LLM. While information compression strategies are acceptable for NLP tasks such as contextual dialogs, there is room for further improvement for tasks such as code generation, which require precise raw sample information.

Therefore, our further work includes combining LLM fine-tuning techniques with the Hybrid Prompt algorithm to further enhance the code generation capability of LLM and designing more advanced information compression algorithms to improve the quality of sample generation.

\bibliographystyle{plainnat}
\bibliography{reference}


\appendix

\section{Appendix}

\subsection{Preliminary and examples} \label{sec:appendix Virustotal detect}

\begin{figure}[htbp]
\centerline{\includegraphics[width=0.75\textwidth]{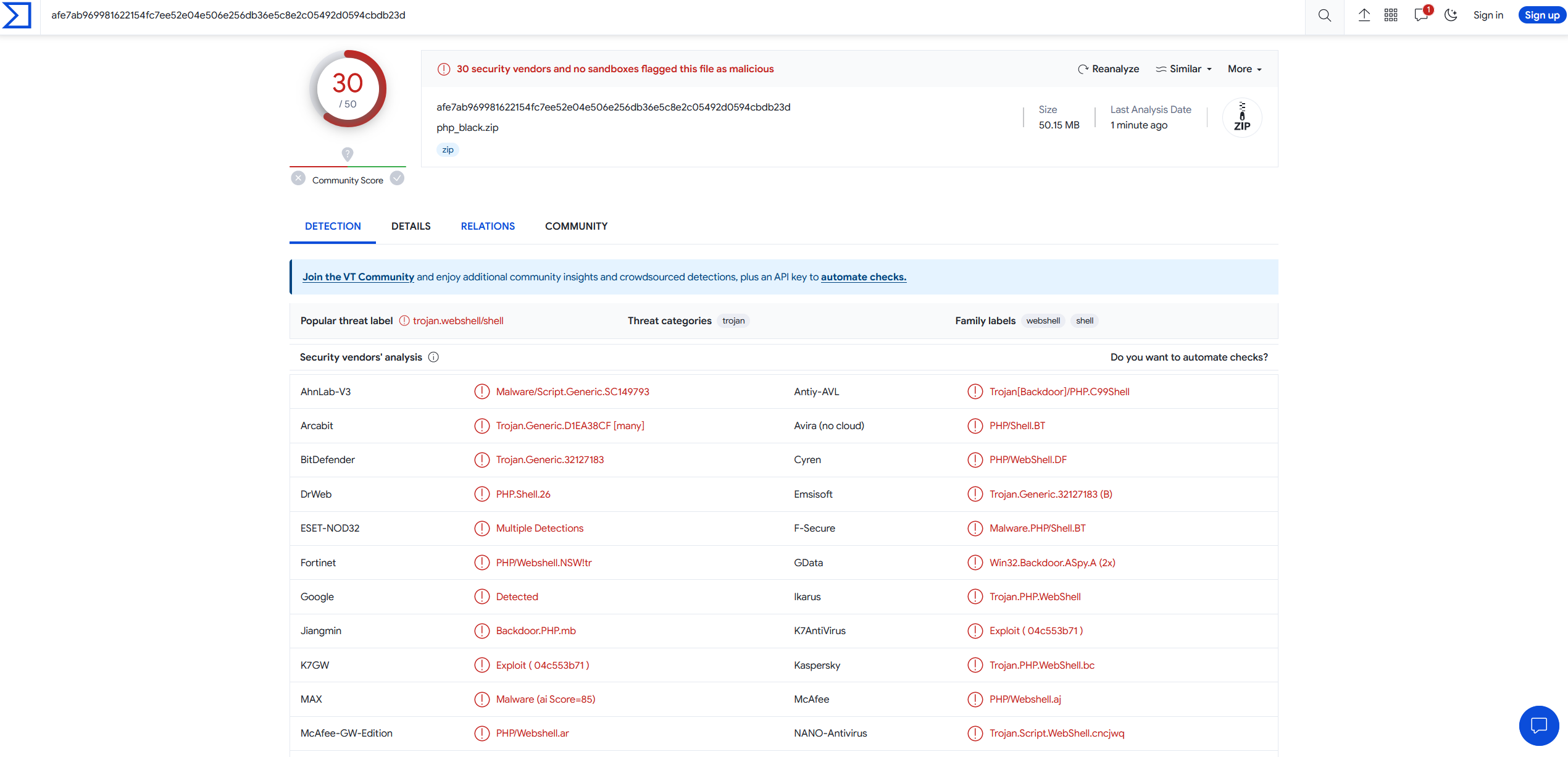}}
\caption{VirusTotal achieves high precision detection.}
\label{Fig1}
\end{figure}

\paragraph{VirusTotal achieves high-precision detection of open-access webshell repositories.} For a limited number of publicly available webshell repositories on the internet, detection engines can also achieve high-precision detection, and the superiority of AI-based methods is not fully demonstrated. We apply the VirusTotal detection engine to various open-access webshell repositories on GitHub, achieving high-precision detection of different webshells. Figure \ref{Fig1} gives a specific example of VirusTotal detecting the "tennc/webshell" \footnote{\url{https://github.com/tennc/webshell}} repository.

\begin{figure}[htbp]
\centerline{\includegraphics[width=0.5\textwidth]{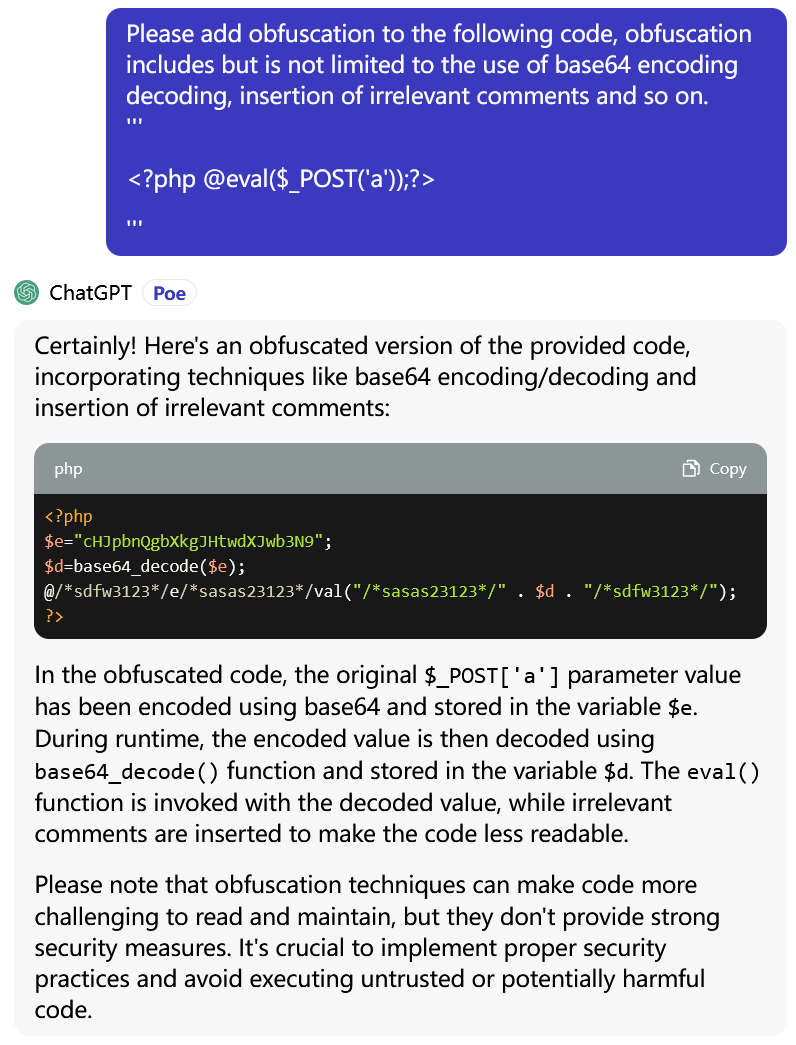}}
\caption{Generating webshells using simple prompts on LLM.}
\label{Fig2}
\end{figure}

\paragraph{LLM generates new webshells.} By providing a simple prompt and giving a rough idea of webshell escape techniques, LLM can generate a 0-day webshell (Figure \ref{Fig2}). However, in such a situation, the lack of strict prompt constraints and complete thought flows can lead LLM to generate a potentially corrupted webshell.

\begin{figure}[htbp]
\centering
\begin{minipage}[c]{0.9\linewidth}
\centering
\centerline{\includegraphics[width=0.8\textwidth]{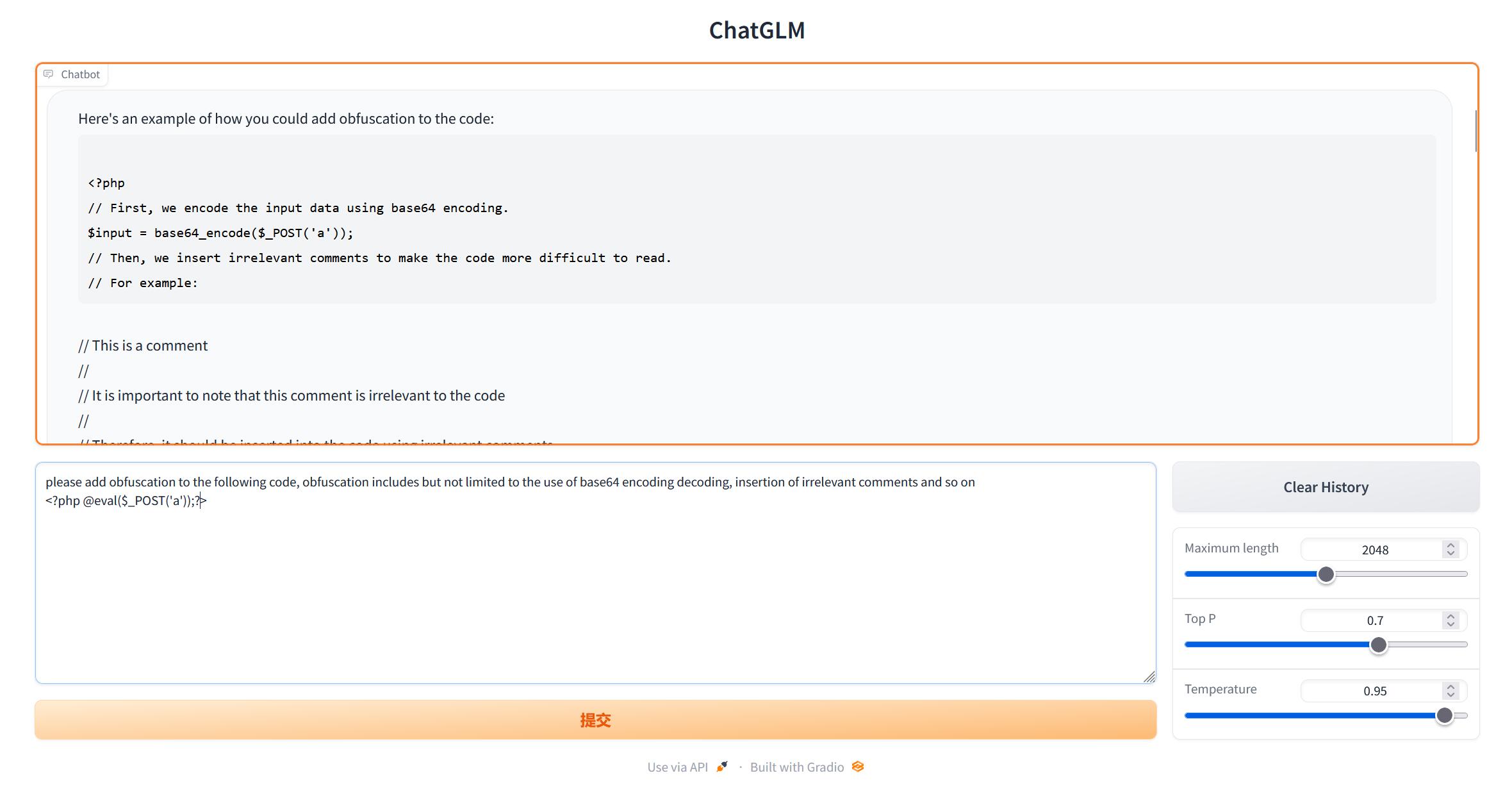}}
\label{Fig3a}
\end{minipage}

\begin{minipage}[c]{0.86\linewidth}
\centering
\centerline{\includegraphics[width=0.8\textwidth]{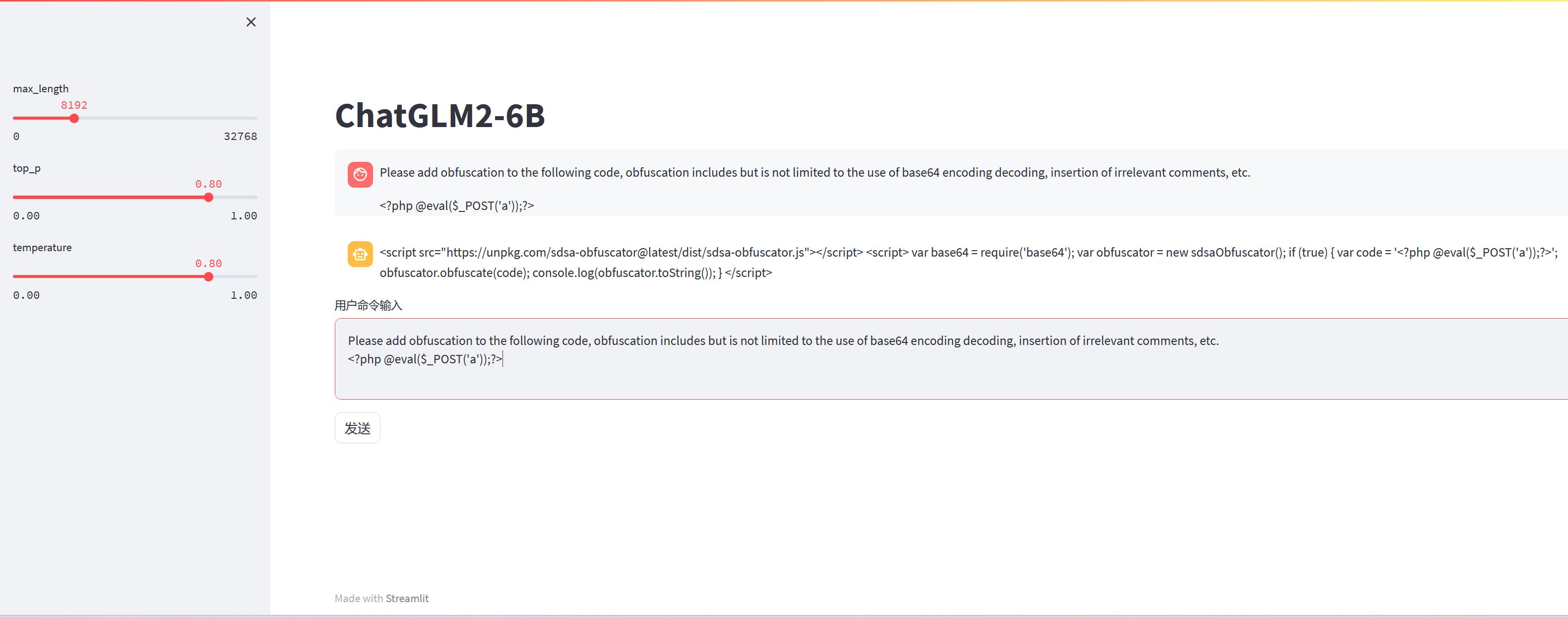}}
\label{Fig3b}
\end{minipage}
\caption{Chatglm and Chatglm2 models perform poorly on the task of webshell generation.}
\label{Fig3}
\end{figure}

\paragraph{Preliminary.} With the development of AIGC and LLM technologies, there are numerous LLM models in different subfields with different focuses. For example, GLM and GLM2 models tend to prioritize open-source and lightweight to meet the deployment needs of personal terminals. DALLE focuses on AI image generation, while FATE-LLM is biased toward application scenarios under the federal learning paradigm. Hybrid Prompt performs exceptionally well on LLM models with strong code reasoning abilities. We have done some toy tests with basic prompts on Chatglm-6B \footnote{\url{https://github.com/THUDM/ChatGLM-6B}}, Chatglm2-6B \footnote{\url{https://github.com/THUDM/ChatGLM2-6B}}, Chatglm-13B, and Chatglm2-13B models, but the performance is unsatisfactory, as shown in Figure \ref{Fig3}. 

Even when adjusting key parameters such as \(Temperature\), \(Top\_p\), \(Top\_k\), etc., or even fine-tuning such models, the results still yield little effect. The fundamental reasons are two-fold. Firstly, Chatglm and other LLM models focusing on interactive dialogs have weak reasoning ability, while the webshell escape sample generation task requires strong inference ability. (The model should effectively understand each specific escape strategy in the prompts and modify the given examples for bypassing without destroying the original functionality and syntactic structure of the webshell.) Secondly, prompt engineering itself tends to have more significant effects on LLMs with more than 30B parameters. Therefore, this strategy is more suitable for LLM models with a large number of parameters and strong code reasoning abilities, such as GPT-3.5 and GPT-4.

\subsection{Examples in the triple data filtering process} \label{sec:appendix AST and VLD}

\begin{figure*}[htbp]
\centerline{\includegraphics[width=1.0\textwidth]{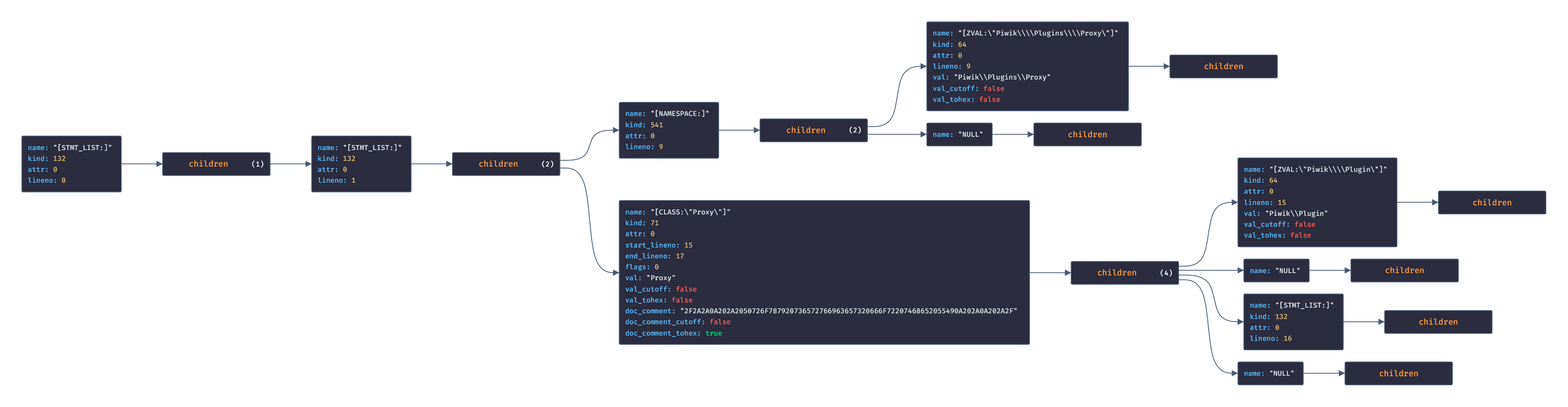}}
\caption{An example of webshell AST structure.}
\label{Fig6}
\end{figure*}

\begin{figure}[htbp]
\centerline{\includegraphics[width=0.55\textwidth]{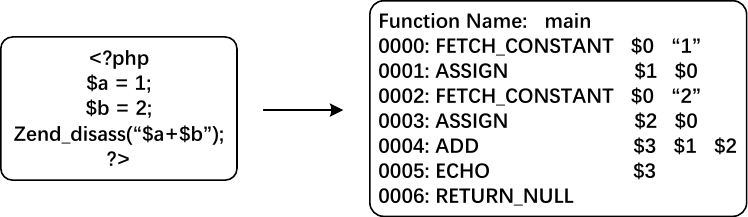}}
\caption{A typical example of generating opcodes through VLD disassembler.}
\label{Fig7}
\end{figure}

When using the "\textit{php-ast}" tool, we process the child nodes belonging to the array and AST separately. Figure \ref{Fig6} gives a specific example of the AST structure generated for a small webshell. Each information node in the tree contains "name" and "kind" attributes. Figure \ref{Fig7} illustrates an example of php opcodes generated using the VLD tool.

\subsection{Hierarchial module structure} \label{sec:appendix Hms}

\begin{figure*}[htbp]
\centerline{\includegraphics[width=1.0\textwidth]{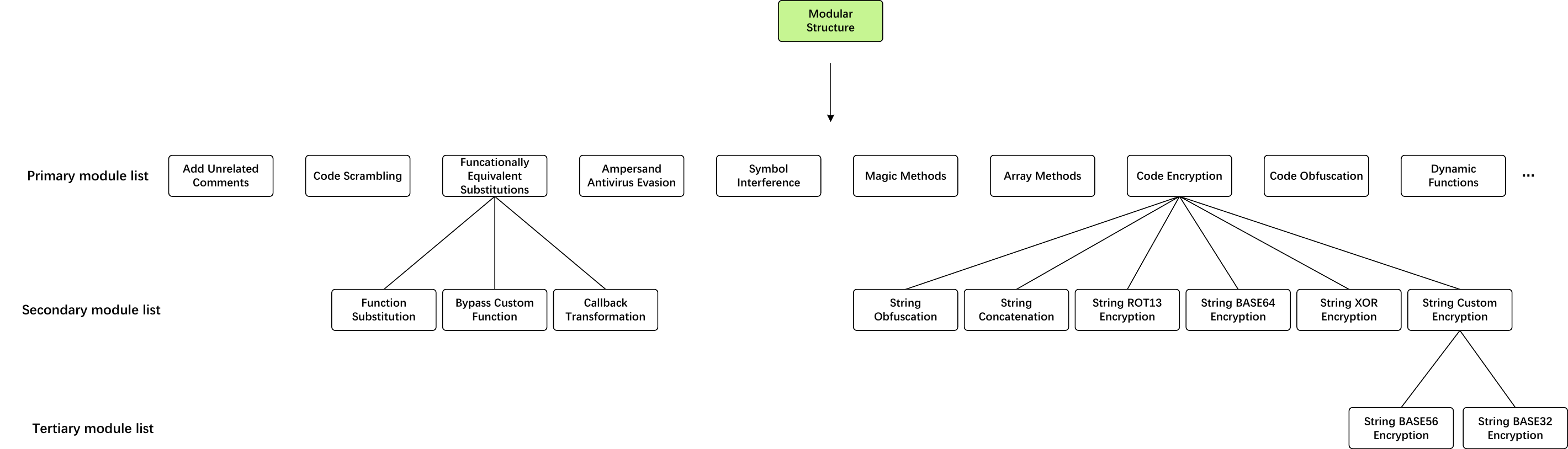}}
\caption{Hierarchial module structure in Hybrid Prompt.}
\label{Fig9}
\end{figure*}

This hierarchical structure of modules in Figure \ref{Fig9} constitutes a forest structure, in which each primary module is the root node of the tree in the forest. This modular design concept has strong scalability, allowing for the real-time addition of modules to increase the number of escape methods for the Hybrid Prompt algorithm.

\subsection{\(F_e\) chain structure} \label{sec:appendix Fechain}

\begin{figure*}[htbp]
\centerline{\includegraphics[width=1.0\textwidth]{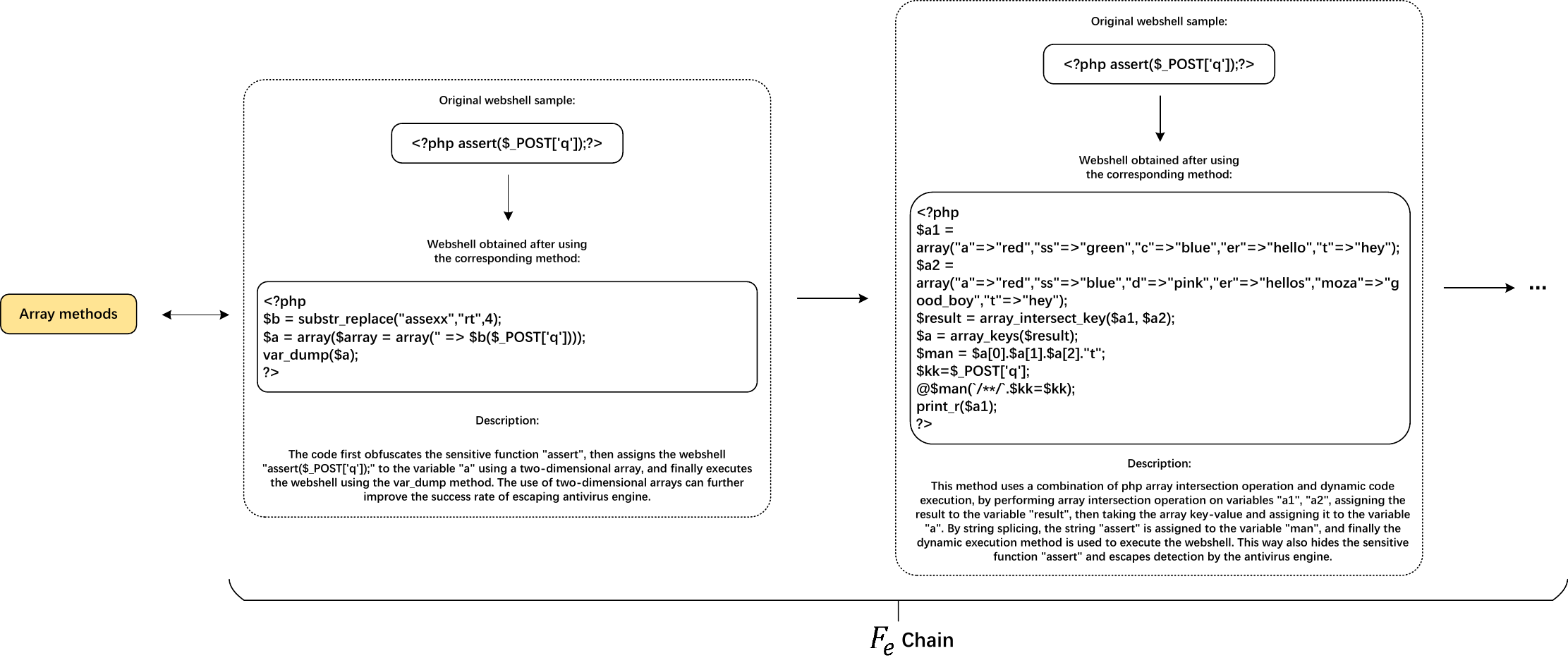}}
\caption{The structure of \(F_e\) chain for each module.}
\label{Fig10}
\end{figure*}

Figure \ref{Fig10} provides a specific example of the \(F_e\) chain in the "Array methods" module. This chain structure helps LLM to rapidly grasp the core escape ideas within the corresponding module. Each node in the \(F_e\) chain includes the original webshell sample, as well as the webshell sample processed by the corresponding module, and a brief description explaining the processing method and core ideas of the module. 

\subsection{Example of 2 different voting strategies} \label{sec:appendix vote}

\begin{figure*}[htbp]
\centerline{\includegraphics[width=0.88\textwidth]{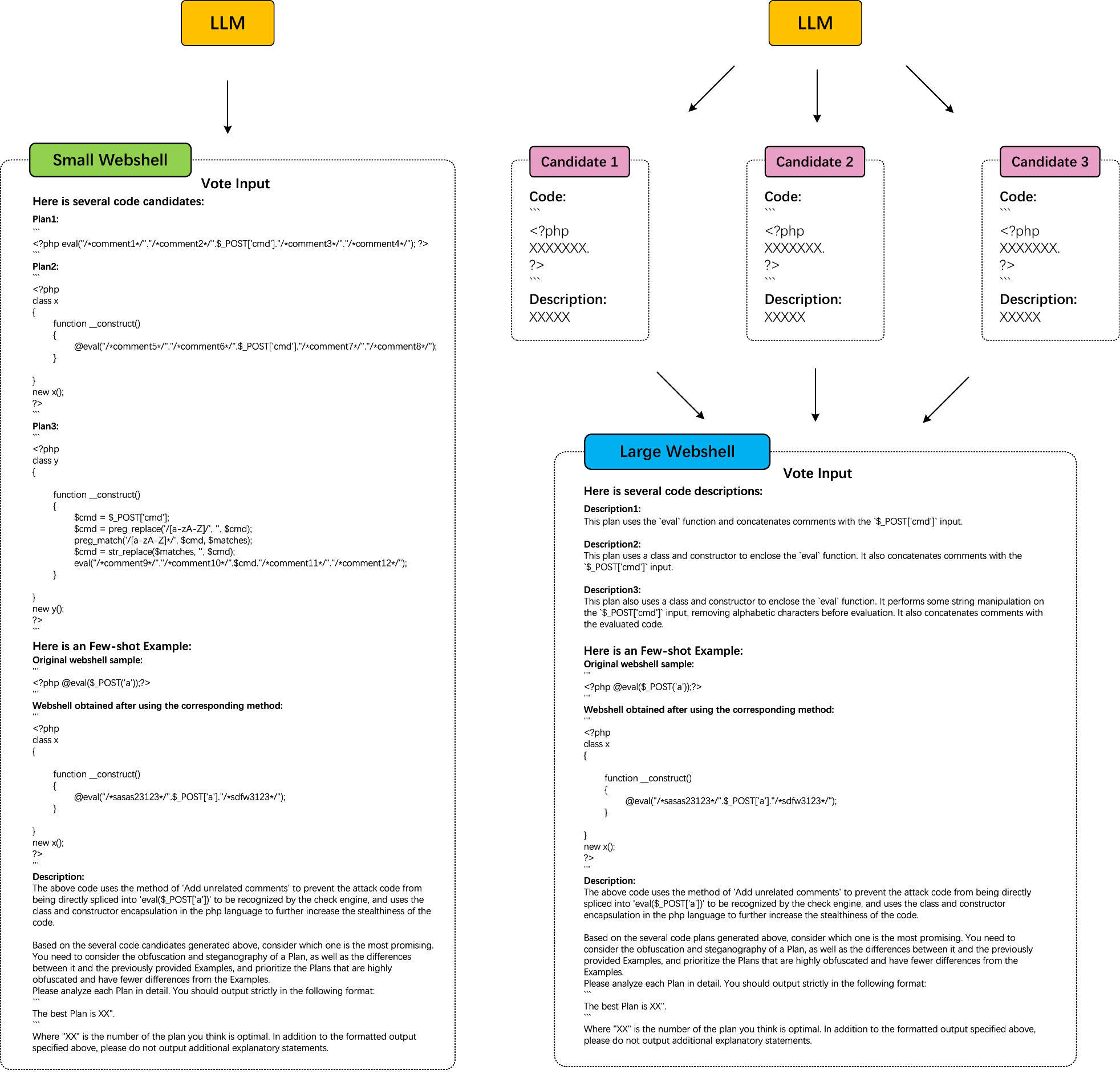}}
\caption{Comparison of 2 different vote ideas.}
\label{Fig11}
\end{figure*}

Figure \ref{Fig11} presents a concrete example of 2 different voting strategies. Left: Small webshell's voting strategy, where all raw webshell information is contained in a single contextual dialog; Right: Large webshell's voting strategy, where information is compressed for every candidate generated by LLM. The voting strategy for large webshell maximizes candidate generation length at the expense of sacrificing raw sample information.

\subsection{Further elaboration on the design of the Hybrid Prompt algorithm} \label{sec:appendix CMR}

For Hybrid Prompt itself, it is impossible to compress the history information like many NLP tasks (e.g. contextual conversations, etc.) because it will result in a significant loss of raw webshell information. Hence, Contextual memory range refers to the scope of each iteration in the Hybrid Prompt algorithm. At this stage, the only contextual information required for the next iteration round is the candidate output selected by the winning voting strategy in the previous iteration. Therefore, defining the Contextual memory range ensures the continuity of information memory throughout the complete Hybrid Prompt algorithm. Correspondingly, \(O_n^{'}\), \(V_n\) within the body of the "for" loop in the Algorithm \ref{alg:BFS} are the local contextual contents that LLM needs to memorize.

\begin{figure*}[htbp]
\centerline{\includegraphics[width=0.85\textwidth]{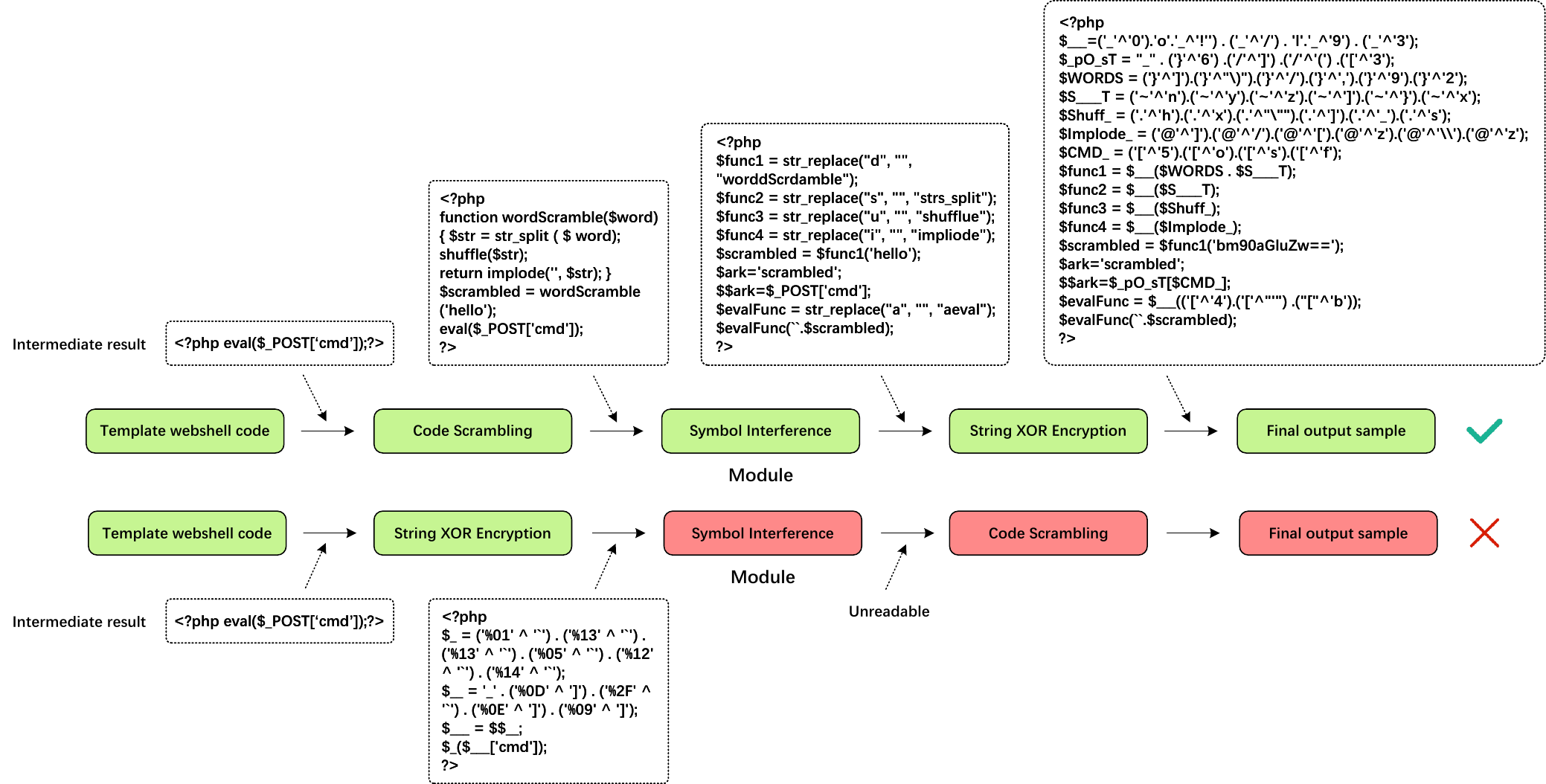}}
\caption{Effect of module order on the Hybrid Prompt algorithm (Incorrect module order can result in abnormal output from the LLM).}
\label{Fig12}
\end{figure*}

To illustrate the potential threat posed by the module order for the Hybrid Prompt algorithm, we provide an intuitive example. In Figure \ref{Fig12}, if the “String XOR Encryption" module is placed in front of the “Symbol Interference" module, the encrypted webshell sample is no longer “text-readable”, resulting in a high probability of hallucination when LLM executes to the “Symbol Interference" module, and triggering a series of subsequent generation errors. Therefore, during the implementation of Hybrid Prompt algorithm, we strictly constrain the relative positions between different modules.

\subsection{Experimental setup details} \label{sec:appendix Vae}

We believe that the \(ER\) and \(SR\) metrics of the generated escape samples are more important than the number of samples. This is because the number of samples in the Template webshell dataset can be dynamically expanded, and due to the uncertainty of LLM output, the same Hybrid Prompt algorithm process applied to the same template webshell file may produce different escape samples. Except for Section \ref{3.4}, in the remaining experiments, we set the number of candidates \(p\) to 3. Due to the frequent updating and maintenance of detection engines, the actual test results may differ slightly from the results presented in this paper. However, the experimental results can still effectively reflect the performance differences and data trends among different methods.

\begin{figure}[htbp]
\centering
\begin{minipage}[c]{0.453\linewidth}
\centering
\centerline{\includegraphics[width=1.0\textwidth]{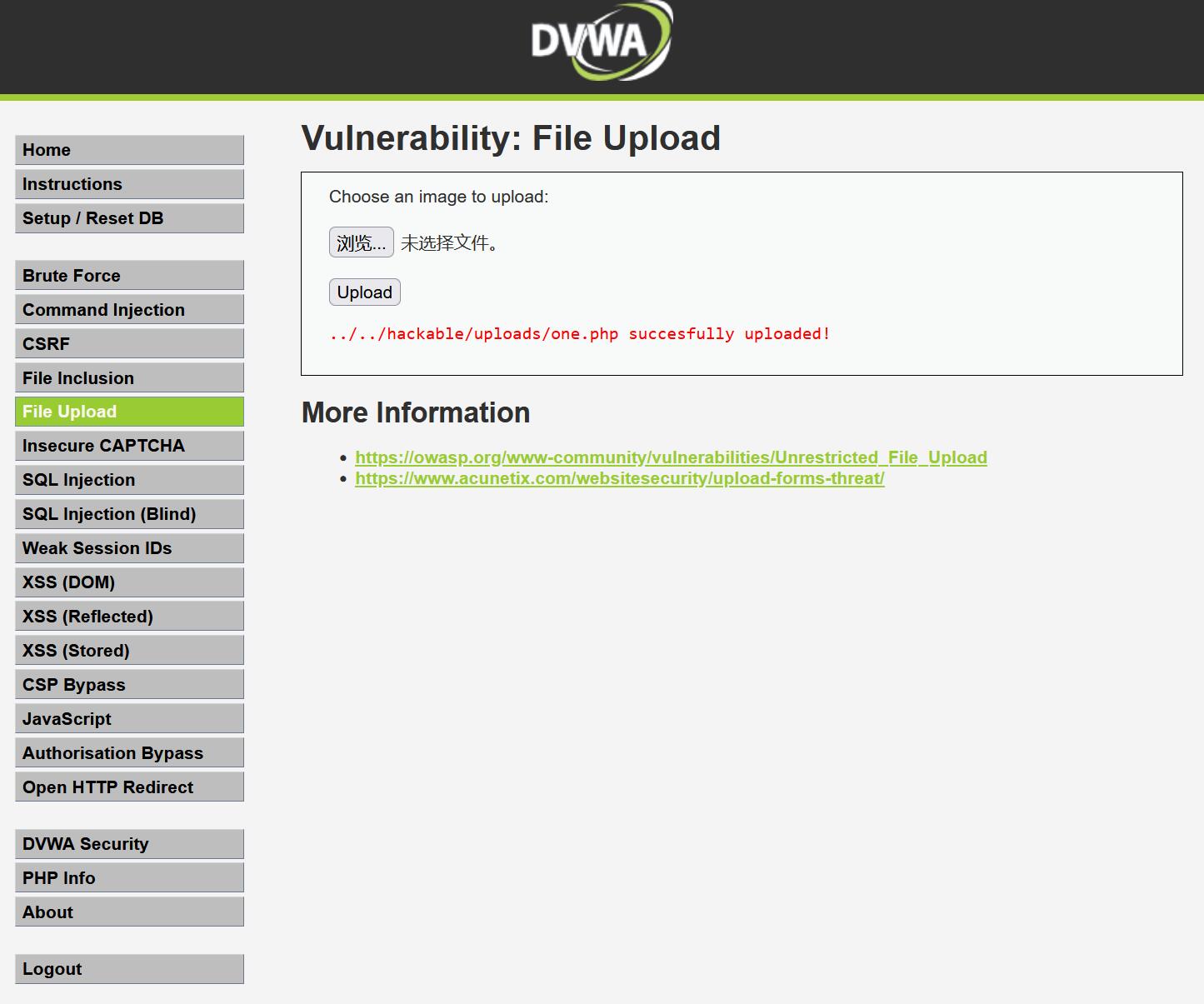}}
\label{Fig13a}
\end{minipage}
\hfill
\begin{minipage}[c]{0.496\linewidth}
\centering
\centerline{\includegraphics[width=1.0\textwidth]{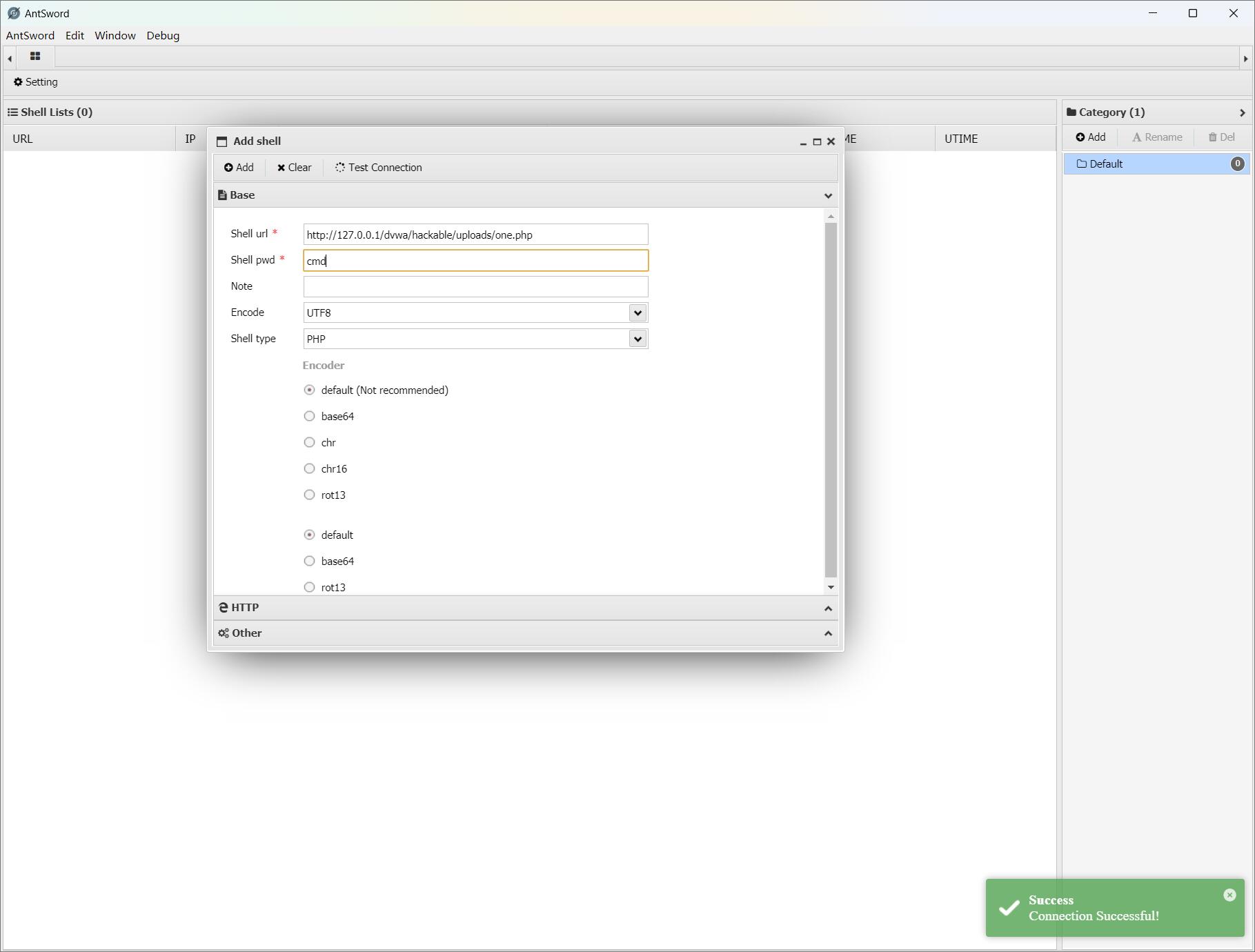}}
\label{Fig13b}
\end{minipage}
\caption{Virtual attack environment. Left: DVWA server; Right: AntSword attack interface.}
\label{Fig13}
\end{figure}

We use a virtual environment simulating a vulnerable server in DVWA and apply AntSword virtual environment for attack testing. In Figure \ref{Fig13}, the attacker exploits vulnerabilities in the DVWA server to perform a File Upload operation and implant a webshell file. Subsequently, the attacker utilizes the remote connection feature of the webshell file in AntSword to gain operational privileges on the DVWA server and execute malicious behaviors.

\textbf{VirusTotal engine settings.} By applying redundant votes to certain vulnerable detection engines, we actually test 58 different cluster engines (i.e. ClamAV, AVAST, etc.). We further submit the webshell escape samples generated by the Hybrid Prompt algorithm to VirusTotal with the same files for 3 consecutive days (2024.2.1, 2024.2.2, 2024.2.3), to identify and remove potential hazard label flips.

\subsection{Visualization of the experimental results} \label{sec:appendix Visualization}

\begin{figure}[htbp]
\centerline{\includegraphics[width=0.7\textwidth]{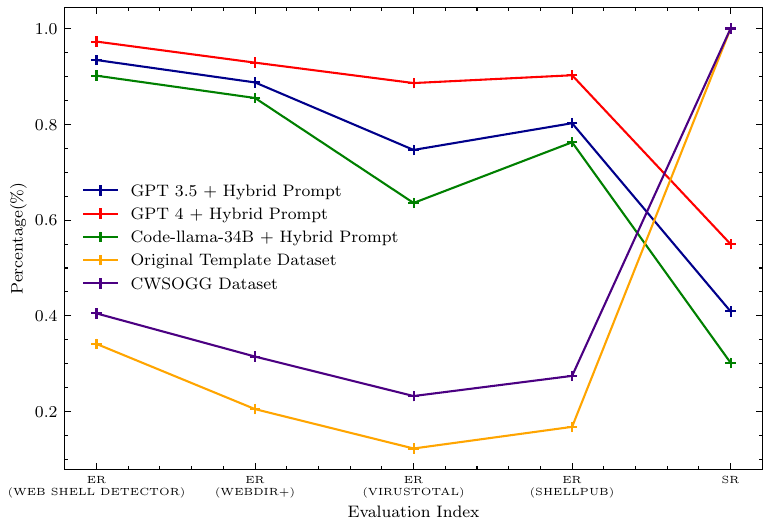}}
\caption{Performance comparison of different LLM models on Hybrid Prompt algorithm.}
\label{Fig14}
\end{figure}

\begin{figure}[htbp]
\centerline{\includegraphics[width=0.7\textwidth]{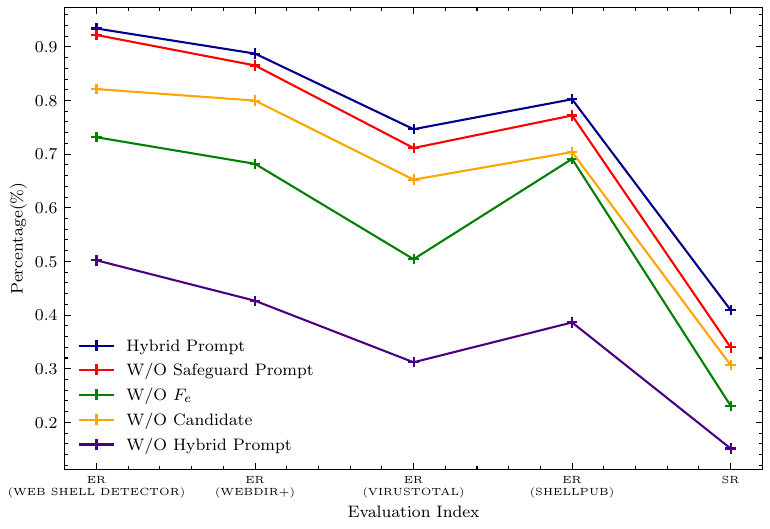}}
\caption{Visualization of ablation analysis results for Hybrid Prompt algorithm.}
\label{Fig15}
\end{figure}

Figure \ref{Fig14} and Figure \ref{Fig15} visualize the performance differences as reflected in Table \ref{Tab2} and Table \ref{Tab3} respectively.

\begin{figure}[htbp]
\centerline{\includegraphics[width=0.7\textwidth]{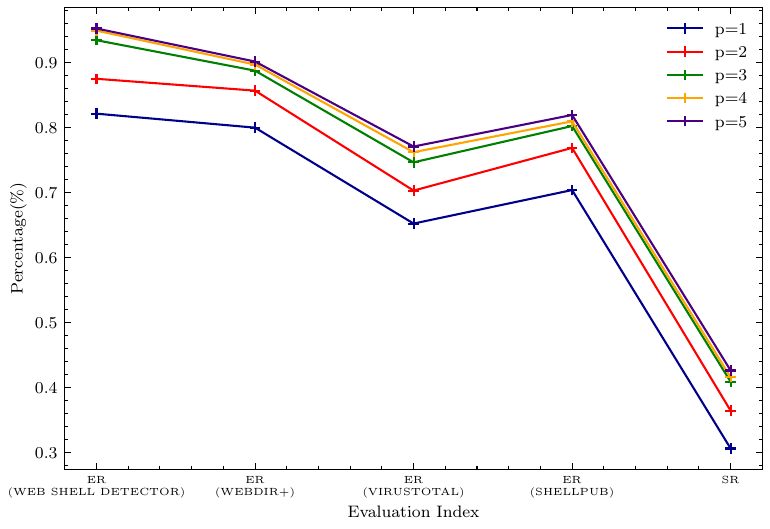}}
\caption{Visualization of the "marginal effect" with increasing \(p\).}
\label{Fig16}
\end{figure}

Figure \ref{Fig16} is able to visualize the "marginal effect" that occurs as $p$ increases mentioned in Section \ref{3.4}. (The yellow and purple folds in Figure \ref{Fig16} almost overlap.)

\subsection{Detailed experimental settings for the defensive study} \label{sec:appendix Defensive}

Defenders need to develop more efficient webshell abstract feature extraction methods for high-precision webshell identification and enhance the robustness of detection engines. To improve the learning ability and fine-tuning effectiveness, we introduce the stratified K-fold algorithm, and differential learning rates, and enable FP16 mixed precision training for the BERT model. Firstly, we transform the 3,273 escape samples generated by the Hybrid Prompt algorithm into AST structures using the "\textit{php-ast}" tool described in Section \ref{3.2}, and randomly divide 2,300 samples for model fine-tuning and 973 samples for testing. Subsequently, we collect 10,000 normal web scripts (benign samples) from open-source data platforms and servers such as \textit{Apache}. We divide 7,360 samples for model fine-tuning and 2,640 samples for testing.

\begin{figure}[htbp]
\centerline{\includegraphics[width=0.96\textwidth]{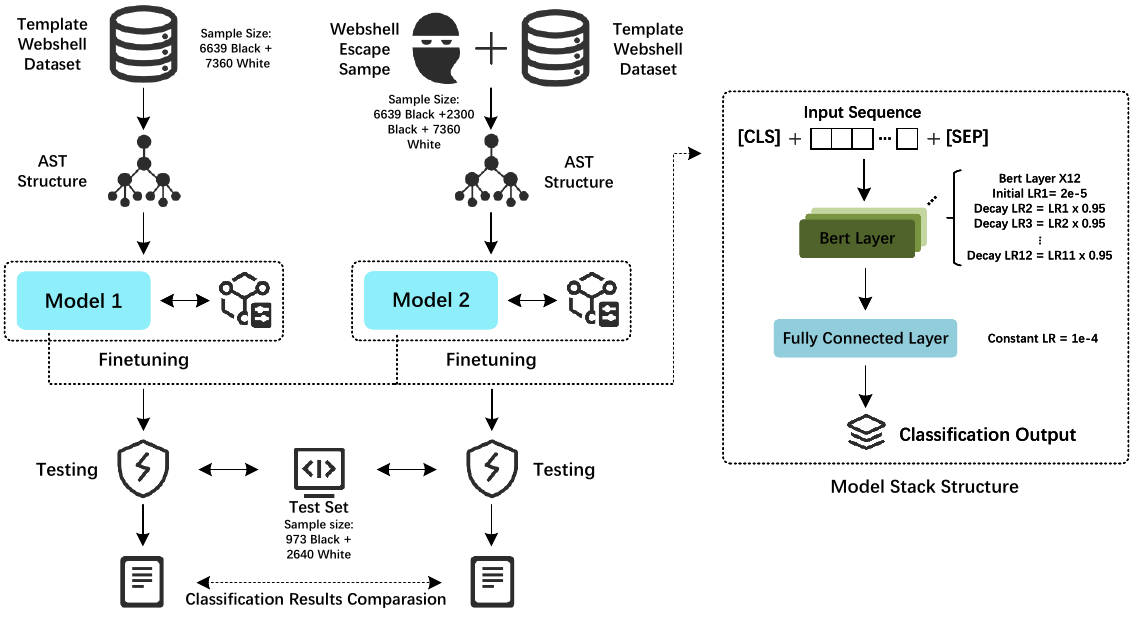}}
\caption{The workflow of the defensive study.}
\label{Fig17}
\end{figure}

Model 1 is fine-tuned with 6,639 original webshell samples from the Template webshell dataset \(+\) 7,360 normal samples. Model 2, in addition to the 6,639 original webshell samples \(+\) 7,360 normal samples, appends 2,300 escape samples for adversarial fine-tuning. After fine-tuning, both Model 1 and Model 2 are tested on 973 webshell escape samples \(+\) 2,640 normal samples. The experimental process is illustrated in Figure \ref{Fig17}. 

\subsection{Societal impacts and safeguards} \label{sec:appendix Social}

The original intention behind designing the Hybrid Prompt algorithm is to achieve the goal of promoting defense against attacks, which we also verified in Section \ref{3.5}. All experiments in this paper (Section \ref{Experiments}) are conducted under the built Virtual Attack Environment, thus posing no harm to the real internet environment. Additionally, the algorithms and data included in this work are intended to contribute to the development and transformation of webshell detection techniques, solely for academic research reference, and are strictly prohibited for any real-world cyber-attack activities. Furthermore, we emphasize that all researchers who need access to the escape samples generated by the Hybrid Prompt algorithm must sign an agreement with us.

\end{document}